\documentclass[amsmath,amssymb,pra,aps,twocolumn,reprint]{revtex4-2}

\usepackage{graphicx}
\usepackage{dcolumn}
\usepackage{bm}
\usepackage[colorlinks=true, allcolors=blue]{hyperref}
\usepackage{slashed}
\usepackage{multirow}

\begin{document}

\title{The blackbody radiation vibrational level shifts in the ground electronic state of N$_{2}^{+}$}
\author{ T. Zalialiutdinov$^{\,1,\,2}$}
\email[E-mail:]{t.zalialiutdinov@spbu.ru}
\author{Y. Demidov$^{\,2,\,3}$}
\author{D. Solovyev$^{\,1,\,2}$}
\affiliation{ 
$^1$ Department of Physics, St. Petersburg State University, Petrodvorets, Oulianovskaya 1, 198504, St. Petersburg, Russia\\
$^2$ Petersburg Nuclear Physics Institute named by B.P. Konstantinov of National Research Centre 'Kurchatov Institut', St. Petersburg, Gatchina 188300, Russia\\
$^3$ 
Physics Department, Saint Petersburg Electrotechnical University ''LETI'', Prof. Popov Str. 5, Saint Petersburg 197376, Russia
}

\date{\today}

\begin{abstract}
Homonuclear molecules have emerged as a crucial component in the pursuit of frequency standards, offering a promising avenue for the discovery of new physics phenomena that transcend the standard model. They also provide a unique approach to constraining variations in fundamental constants over time, thereby complementing the capabilities of atomic clocks. A notable challenge faced by molecular and single atomic quantum systems is the management of blackbody radiation (BBR), which introduces significant systematic errors and is challenging to regulate effectively. To address this issue, we perform {\it ab-initio} quantum chemical calculations to accurately determine the potential energy curve and the polarizability tensor for the ground state of the N$_2^+$ molecular ion, one of the most promising candidates for searching for variation of $m_{e}/m_{p}$ and creating frequency standards. We then calculate the BBR shifts affecting the vibrational levels of the ground electronic X$^2\Sigma_g^+$ state, marking a substantial contribution towards the precise experimental measurements. 
\end{abstract}

\maketitle
\section{Introduction}

Molecular nitrogen ions are excellent candidates for investigating physics beyond the Standard Model of particle physics and cosmology \cite{Tiberi2024}. Their transitions exhibit sensitivity to variations in the electron-to-proton mass ratio, $m_e/m_p$, while maintaining remarkable stability against environmental fluctuations \cite{Dirac_main,Kajita2014}. Measurements of the vibrational transitions of nitrogen ions promise to achieve frequency uncertainties better than $10^{-20}$, surpassing, the accuracy of the best atomic clocks currently available  \cite{Barontini2022,PhysRevLett.97.233001}.

Initial interest in this system was piqued by the dissociative photoionization of molecular nitrogen, which plays a pivotal role in various environments, including Earth's upper atmosphere (ionosphere), planetary atmospheres, astrophysical media (interstellar medium), and plasma \cite{LieSvendsen1992, Dutuit2013}. Recently, this molecule has also been proposed as a candidate for ultra-precise molecular clocks and as a tool for studying variations in fundamental constants, such as the electron-to-proton mass ratio, $m_e/m_p$ \cite{Kajita2014}.

The N$_{2}^{+}$ molecule plays a significant role in various fields of physics. For instance, related studies on airglow and planetary entry physics have been discussed in the literature, see, e.g., \cite{Qin2017}. Recently, it has also been observed in a few comets, most notably in Comet C/2016 R2, which exhibited unusually bright emission lines due to N$_{2}^{+}$ \cite{Biver2018}. Numerous experimental and theoretical studies have aimed to accurately determine vibrational energy levels and rotational constants, from which rovibrational energy levels and, consequently, transition frequencies can be readily calculated \cite{Ayari2020,Ferchichi2022}. However, despite extensive calculations of various spectroscopic properties, dipole polarizability functions and thermal Stark shifts have received comparatively less attention.

As a homonuclear diatomic molecule, N$_2^+$ has no electric dipole (E1) transitions between different rovibrational states in its X$^2\Sigma_g^+$ electronic ground state, making effects induced by external fields, such as blackbody radiation, expected to be negligible \cite{Gilmore_1992,Kajita2014}. However, to the best of our knowledge, the exact values of these shifts remain unknown.

The paper is organized as follows. In Section~\ref{part1}, we present the theoretical basis for calculating thermal Stark shifts in many-electron systems (diatomic molecules). The results, numerical methods, and details of the electronic structure calculations for N$_2$ and N$_2^+$ molecules, as well as spectroscopic constants within the coupled cluster approach, are discussed in Section~\ref{part0}, where the validity of the chosen approximations is also addressed. The quantum chemical calculations were performed using the \texttt{CFOUR 2.1}, \texttt{NWChem 7.2.2}, and \texttt{DIRAC24} software programs \cite{cfour, nwchem, DIRAC24}. Calculations of the vibrational structure and the corresponding thermal Stark shifts are presented in Section~\ref{results}. Final conclusions are provided in the last part of the paper.

\section{Blackbody Radiation Shifts in diatomic molecules}
\label{part1}

The calculation of thermal shifts in atoms and molecules demands an understanding of both static and dynamic polarizabilities and their dependence on internuclear distance. These dynamic polarizabilities are obtained through linear response (LR) equations, which significantly increase computational costs compared to binding energy calculations. In addition to determining polarizabilities, we must also establish the potential energy curves by solving the nuclear problem. Furthermore, vibrational wave functions for all nonrotating vibrational states of N$_{2}^{+}$ in its electronic ground state are needed. This can be reduced to the one-dimensional Schr\"{o}dinger equation for the known interatomic potential $V(R)$ \cite{Yurchenko_2016}:
\begin{eqnarray}
\label{1}
\left(-\frac{\hbar^2}{2\mu} \frac{d^2 }{dR^2} + \frac{J(J+1)}{2\mu R^2}+  V(R)\right) \psi_{\nu J}(R) 
\\\nonumber
= E_{\nu J} \psi_{\nu J}(R),    
\end{eqnarray}
where $\psi_{\nu J}$ is the wave function corresponding to the state with vibrational and rotational quantum numbers $\nu$ and $J$, respectively. For charged diatomic molecules, the “charge-modified reduced mass”, $\mu$ in Eq.~(\ref{1}), is defined as follows \cite{Watson1980}:
\begin{eqnarray}
    \mu=\frac{m_{1}m_{2}}{m_{1}+m_{2}-m_{e}q},
\end{eqnarray}
where $m_1$, $m_2$ are the atomic masses, $m_e$ is the mass of an electron and $q$ is the net charge of the molecule. In general, the second-order differential equation (\ref{1}) is solved numerically using a mesh of radial points, see, for example, \cite{Stol1, Stol2}. Within our implementation, the $V(R)$ function is interpolated using the cubic splines. The finite-difference method is used to perform integration. To validate our results, we compare the numerically computed eigenvalues and eigenvectors of Eq.~(\ref{1}) with those obtained with \texttt{DUO} package \cite{Yurchenko2016}, a specialized software tailored for calculating the vibrational structure of diatomic molecules. The comparison yielded identical results to within two decimal places, confirming the accuracy of our implementation.

The potential curve in Eq.~(\ref{1}) can also be obtained from experimental data by analyzing the vibration spectrum. The Rydberg–Klein–Rees (RKR) method \cite{Singh1962}, widely employed in the analysis of rotational-vibrational spectra of diatomic molecules, enables the reconstruction of potential energy curves from experimentally determined line positions and molecular constants. The commonly used RKR curve relevant to the ground state X$^2\Sigma_g^+$ of the N$^+_{2}$ molecule is given in the works \cite{Singh1966, Lofthus1977}. We, however, performed our own RKR restoration of the molecular potential based on more relevant experimental data from the work \cite{Klynning1982} (see Table 2 therein). The resulting PEC is presented Fig.~\ref{fig:cbs}, while the corresponding vibrational energy levels are given in the last column of Table~\ref{tab:vibr}. It is noteworthy that the empirically obtained vibrational spectrum exhibits excellent agreement, with a relative uncertainty of approximately 0.04\%, when compared to recent calculations conducted using the explicitly correlated multi-reference configuration interaction (MRCI-F12) approach \cite{Ferchichi2022} (see the fourth column in Table~\ref{tab:vibr}). As will be demonstrated below our present calculations within the coupled clusters approach exhibit a slightly lower accuracy, resulting in a relative uncertainty of approximately 0.5\%. However, for all subsequent calculations involving integrals with $\psi_{\nu}(R)$ and polarizabilities as a function of internuclear distance, we choose to utilize the numerical solution of Eq.~(\ref{1}) based on the updated experimental RKR curve rather than the theoretically computed one, as it offers superior accuracy.

\begin{widetext}
\begin{table*}[ht]
\caption{Vibrational energy levels $\nu$ of the ground state X$^2\Sigma_{g}^{+}$ for the molecular ion N$_2^+$ compared with experimental data (last column) and other theoretical results. The theoretical values in the 6th column are obtained at the CCSD(T)+DPT2 level and the CBS limit within \texttt{CFOUR} code. The corresponding fully relativistic results with \texttt{DIRAC} code are given in the 7th column. All the values are given in cm$^{-1}$.}
\begin{tabular}{c c c c c c c c c} 
\hline 
& \multicolumn{2}{c}{Other theoretical results} &  &  \multicolumn{2}{c}{\texttt{CFOUR}}  & \texttt{DIRAC} 
& RKR  & RKR 
\\
$\nu$ &\cite{Langhoff1987} &\cite{Nagy2003} &\cite{Ferchichi2022} & CBS & CBS/DPT2 & CBS & \cite{Singh1966,Lofthus1977} & this work\\
\hline 
0 & 0.0 & 0.0 & 0.0 & 0.0 & 0.0 & 0.0 & 0.0 & 0.0\\ 
1 & 2144.7 & 2244.2 & 2175.62 & 2189.7 & 2193.7 & 2189.3 & 2186.3 & 2174.8\\ 
2 & 4256.1 & 4449.8 & 4318.73 & 4342.8 & 4350.1 & 4346.7 & 4318.1 & 4317.1\\ 
3 & 6333.9 & 6616.8 & 6429.17 & 6482.8 & 6473.8 & 6473.8 & 6436.9 & 6426.8\\ 
4 & 8380.2 & 8745.1 & 8506.64 & 8593.3 & 8570.9 & 8570.90 & 8490.9 & 8503.6\\ 
5 & 10394.6 & 10834.8 & 10551.06 & 10653.6 & 10681.5 & 10637.3 & 10548.6& 10547.3\\ 
6 & 12375.3 & 12885.8 & 12562.02 & 12698.2 & 12748.5 & 12671.4 & 12552.2& 12557.9\\
7 & 14321.6 & 14898.2 & 14539.71 & 14707.9 & 14798.4 & 14672.6 & 14530.1& 14535.0\\ 
8 & 16232.3 & 16871.9 & 16484.00 & 16680.9 & 16835.4 & 16640.3 & 16470.8& 16478.5\\ 
9 & 18106.6 & 18807.0 & - & 18616.2 & 18861.1 & 18574.1 & 18374.6& 18388.2\\ 
10 & - & 20703.3 & - & 20513.9 & 20874.6 & 20474.1 & 20245.5 & 20264.0\\ 
11 & - & 22561.2 & - & 22374.8 & 22872.4 & 22340.3 & 22081.3 & 22105.8\\ 
12 & - & 24380.3 & - & 24200.1 & 24849.9 & 24173.5 & 23881.9 & 23913.2\\ 
13 & - & 26160.7 & - & 25990.8 & 26802.0 & 25974.1 & 25646.6 & 25686.2\\ 
14 & - & 27902.6 & - & 27748.3 & 28723.3 & 27742.6 & 27380.1 & 27424.5\\ 
15 & - & 29605.8 & - & 29473.7 & 30609.1 & 29479.4 & 29455.4 & 29128.0\\ 
\hline 
\end{tabular}
\label{tab:vibr}
\end{table*}
\end{widetext}

To estimate the Stark shift at any given $\omega$ frequency, one should first calculate the molecular interaction-induced dynamic polarizability \cite{Iritani2023}:
\begin{eqnarray}
    \alpha_{ij}^{\mathrm{int}} (\omega, R)
    = \alpha_{ij}(\omega, R) - \alpha^{A}_{ij} (\omega)  
    - \alpha^{B}_{ij} (\omega),
\end{eqnarray}
where $\alpha_{ij}(\omega, R)$ are tensor components of
the total molecular polarizability at internuclear separation $R$ and $\alpha^{A} (\omega) $ and $\alpha^{B} (\omega) $ are the atomic
polarizabilities of atoms $A$ and $B$, respectively, at frequency $\omega$. 
As detailed in \cite{Kajita2014}, we will not account for the thermalization of the rotational states. This leads us to consider only the isotropic $J=0$ states below, for which we take the trace of the polarizability tensor as follows:
\begin{eqnarray}
    \alpha^{\mathrm{int}} (\omega, R)
    = \left[2\alpha_{xx}^{\mathrm{int}}(\omega, R) + \alpha_{zz}^{\mathrm{int}} (\omega, R)\right]/3.
\end{eqnarray}

Once Eq.~(\ref{1}) has been solved and the set of eigenvalues $E_{\nu}$ and eigenfunctions $\psi_{\nu}$ has been obtained, it is necessary to calculate the polarizability of each vibrational level $\nu$ by averaging the electronic polarizability $\alpha^{\mathrm{int}}(\omega,R)$ over the level’s vibrational wave function $\psi_{\nu}(R)$:
\begin{eqnarray}
\label{avg}
    \alpha^{\mathrm{int}}_{\nu}(\omega) = \int\limits_{0}^{\infty}|\psi_{\nu}(R)|^2 \alpha^{\mathrm{int}}(\omega,R)\,dR.
\end{eqnarray}

The final step entails computing the Stark shifts induced by the fluctuating electric field generated by equilibrium thermal radiation, commonly known as black-body radiation (BBR) \cite{Itano}. The corresponding spectral radiance at temperature $T$ (in Kelvin) is
\begin{eqnarray}
    B_{T}(\omega)=\frac{\hbar \omega^3}{4\pi^3 c^2}\frac{1}{\exp(\hbar\omega/k_{B}T)-1},
\end{eqnarray}
where $k_{B}$ is the Boltzmann constant, $c$ is the speed of light and $\hbar$ is the reduced Planck constant. 

In contrast to atomic systems, where thermal effects are primarily evaluated using direct summation over states (SOS) \cite{Safronova_2012, Zalialiutdinov_2017, Zalialiutdinov_2022} and dynamic polarizabilities are determined over a broad frequency range, molecular systems benefit from a different approach. Specifically, it is advantageous to use a Taylor series expansion of the dynamic polarizability at low frequencies. As a result, the leading contribution comes from static polarizability, with dynamic corrections being significantly suppressed by higher powers of temperature. This approximation is generally valid when $k_{B}T \ll \Delta E$, where $\Delta E$ is the transition energy to the nearest state allowed by dipole selection rules \cite{Farley1981}. Consequently, the shift induced by blackbody radiation (BBR) is expressed as a series \cite{Iritani2023}:
\begin{eqnarray}
\label{BBR}
    \Delta f_{\nu}=-\sum\limits_{\substack{n=\\0,\,2,\,4\dots}}\frac{c_{n} \alpha^{(n)}_{\nu}(0)}{4\pi^3\epsilon_{0}c^3}\left( \frac{k_{B}T}{\hbar}\right)^{4+n},
\end{eqnarray}
where $\epsilon_{0}$ is the vacuum permitivity 
and 
\begin{eqnarray}
 \alpha^{(n)}_{\nu}(0)=\frac{d^{n} \alpha_{\nu}(\omega)}{d\omega^{n}}\Bigg|_{\omega=0}.
\end{eqnarray}
The coefficients $c_{n}$ in Eq. (\ref{BBR}) are defined as follows
\begin{eqnarray}
\label{cn}
c_{n}=\int_{0}^{\infty}x^{3+n}/(e^{x}-1)dx.
\end{eqnarray}  
The integral in Eq. (\ref{cn}) can be taken analytically, giving the values
\begin{eqnarray}
c_{0}=\frac{\pi ^4}{15}, \quad c_{2}= \frac{8\pi ^6}{63}.
\end{eqnarray}

\section{Electronic structure calculations}
\label{part0}

Performing all the necessary steps to estimate the Stark shift, we start by constructing the electronic structure of the molecular ion N$^+_2$. Numerical calculations of the nonrelativistic potential curve $V(R)$ for the ground state were carried out within \texttt{CFOUR} code at CCSD(T) level of treatment using Dunning's correlation-consistent basis sets,  aug-cc-PV$n$Z ($n=T,\,Q$). The results are listed in Table \ref{tab:curve}.
\begin{widetext}
\begin{table*}[ht]
\caption{Potential energy of the ground X$^2\Sigma_{g}^{+}$ state of the N$_2^+$ molecular ion. Calculations were performed at the CCSD(T) level of theory using Dunning's correlation-consistent basis sets, aug-cc-PV$n$Z ($n=$ T, Q), with the \texttt{CFOUR} and \texttt{DIRAC} codes. The nonrelativistic results, including scalar relativistic corrections within second-order direct perturbation theory (DPT2), are presented in the 5th column \cite{Stopkowicz_2018} (only the CBS limit is given for brevity). Fully relativistic calculations using the \texttt{DIRAC} code are provided in the last three columns. All the values are given in cm$^{-1}$.}
\begin{tabular}{c | c c c c | c c c}

\hline
&\multicolumn{4}{c|}{\texttt{CFOUR}}&\multicolumn{3}{c}{\texttt{DIRAC}}\\
R, a.u. & aug-cc-PVTZ & aug-cc-PVQZ & CBS & CBS/DPT2& aug-cc-PVTZ & aug-cc-PVQZ & CBS\\
\hline
1.6 & -108.439907 & -108.502453 & -108.588367 & -108.631242& -108.544281 & -108.604818 & -108.648994\\
1.7 & -108.641694 & -108.700242 & -108.780665 & -108.827755& -108.702075 & -108.759667 & -108.801693\\
1.8 & -108.742131 & -108.798661 & -108.876312 & -108.924945& -108.802360 & -108.857886 & -108.898404\\
1.9 & -108.801341 & -108.856522 & -108.932319 & -108.981919& -108.861499 & -108.915611 & -108.955098\\
2.0 & -108.831328 & -108.885631 & -108.960224 & -109.010377& -108.891476 & -108.944637 & -108.983429\\
2.1 & -108.840910 & -108.894648 & -108.968464 & -109.018915& -108.901099 & -108.953614 & -108.991937\\
2.2 & -108.836581 & -108.889941 & -108.963239 & -109.013850& -108.896847 & -108.948909 & -108.986900\\
2.3 & -108.823114 & -108.876201 & -108.949123 & -108.999823& -108.883481 & -108.935202 & -108.972944\\
2.4 & -108.804004 & -108.856869 & -108.929486 & -108.980232& -108.864485 & -108.915924 & -108.953460\\
2.5 & -108.781791 & -108.834457 & -108.906800 & -108.957559& -108.842389 & -108.893575 & -108.930927\\
2.6 & -108.758308 & -108.810779 & -108.882855 & -108.933607& -108.819013 & -108.869957 & -108.907132\\
2.7 & -108.734866 & -108.787133 & -108.858929 & -108.909673& -108.795651 & -108.846351 & -108.883349\\
2.8 & -108.712399 & -108.764439 & -108.835923 & -108.886684& -108.773220 & -108.823661 & -108.860470\\
2.9 & -108.691569 & -108.743348 & -108.814472 & -108.865297& -108.752369 & -108.802523 & -108.839122\\
3.0 & -108.672846 & -108.724319 & -108.795023 & -108.845974& -108.733564 & -108.783393 & -108.819755\\
\hline
\end{tabular}
\label{tab:curve}
\end{table*}
\end{widetext}

It should be noted that the nonrelativistic energy values obtained in \texttt{CFOUR} coincide with those obtained in the \texttt{NWChem} program packages. The latter is used in the next section to calculate polarizabilities because of its superior numerical stability in calculating molecular properties in all significant digits presented in the 2nd and 3rd columns of Table~\ref{tab:curve}.

To avoid the incompleteness of the basis sets used, it is necessary to make a conversion to the limit of the complete basis set (CBS). The CBS transition was originally introduced by Helgaker et al.~\cite{Helgaker1997} as
\begin{eqnarray}
    E=E_{\mathrm{CBS}}+\frac{A}{n^3},
\end{eqnarray}
where $A$ is a fit parameter and $n$ is the highest cardinal number of atomic gaussians of the basis used. A year later, in a follow-up to the paper \cite{Halkier1998-zu}  the efficiency of the aforementioned formula was evaluated using large basis sets. These results were then compared with the advanced explicitly correlated calculations, resulting in the following equation for the CBS limit:
\begin{eqnarray}
\label{CBS}
  E_{\mathrm{CBS}} = \frac{E_{n-1} (n-1)^3 - E_{n} n^3}{(n-1)^3 - n^3}. 
\end{eqnarray}
Next we use Eq.~(\ref{CBS}) for all potential curves and will generalize Eq.~(\ref{CBS}) to the polarizability case (see Eq.~(\ref{CBS_alpha})).

In addition to the purely nonrelativistic evaluation in the program packages \texttt{CFOUR} and \texttt{NWChem}, we performed calculations that take into account scalar relativistic effects. Both perturbation theory (the DPT2 approach in \texttt{CFOUR}, as discussed in \cite{Stopkowicz_2018, Matthews_2020}) and fully relativistic calculations with the \texttt{DIRAC} package were used for this purpose. The results can be found in Table \ref{tab:curve} with appropriate notations (CBS/DPT2 for \texttt{CFOUR} and the CBS column in the \texttt{DIRAC} part).

\begin{figure}[h]
    \centering
    \includegraphics[width=0.45\textwidth]{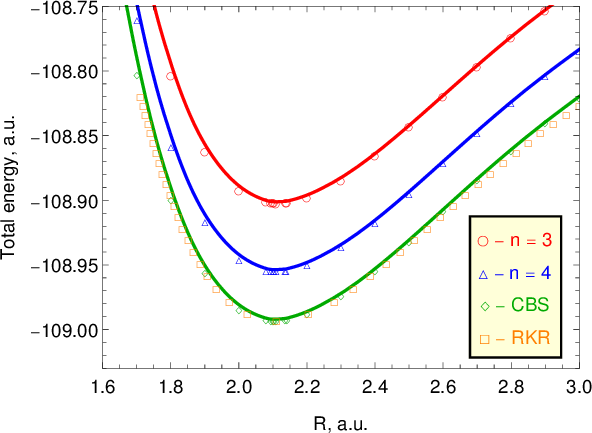}
    \caption{Potential energy curves for the ground X$^2\Sigma_{g}^{+}$ state of the N$_2^+$ molecular ion. Calculations are performed at the CCSD(T) level of theory using Dunning's correlation-consistent basis sets aug-cc-PV$n$Z with $n=T,\,Q$ (triple and quadruple), within the \texttt{DIRAC} code. The curve obtained for the aug-cc-PVTZ basis set is marked by circles, triangles - aug-cc-PVQZ, diamonds - curve for the CBS limit, and squares for the case of determination from experimental data (RKR curve) The experimental RKR curve obtained in the present study is shifted to align with the minimum energy of the CBS curve.} 
    \label{fig:cbs}
\end{figure}

The CBS limit for fully relativistic CCSD(T) results is illustrated in Fig.~\ref{fig:cbs}.
To demonstrate the quality of CCSD(T)/CBS calculations, we compare this potential curve with RKR one. For the visual comparison in Fig.~\ref{fig:cbs} the updated RKR PEC obtained in this work, was shifted to the minimum energy of the CCSD(T)/CBS potential:~$V_{\mathrm{RKR}}(R)-V_{\mathrm{CBS}}(R_{e}^{\mathrm{CBS}})$, where $R_{e}^{\mathrm{CBS}}$ is the equilibrium distance.
It can be seen that the potential curves within the studied range of internuclear distances, from 1.6 to 3.0 a.u., are consistent. 

In recent decades, diatomic systems have often been considered within the context of multireference approaches, including CASSCF (complete active space self-consistent field) and MRCI (multi-reference configuration interaction) \cite{Langhoff1987, Hadjipittas2023, Ferchichi2022}. However, these approaches are significantly more demanding in terms of numerical resources. In particular, the calculation of response properties, such as polarizability, represents a significant challenge. Even in the context of the single-reference coupled clusters, this often requires more computational resources than the calculation of energies.

\begin{figure}[h]
    \centering
    \includegraphics[width=0.45\textwidth]{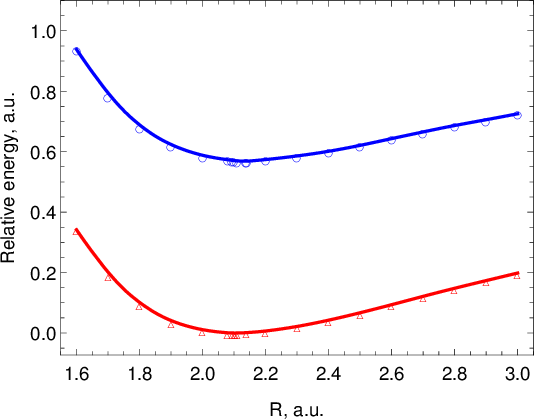}
    \caption{Potential energy curve for the ground state of N$_2^+$ molecule, X$^2\Sigma_{g}^{+}$ (circles) relative to X$^1\Sigma_{g}^{+}$ state of N$_2$ molecule (triangles) calculated by the CCSD(T)/CBS method within \texttt{DIRAC} code.}
    \label{fig:IP}
\end{figure}

The fully relativistic CCSD(T)/CBS method can be realized for the neutral molecule $\rm N_2$ as well. Potential energy curves (PECs) for the ground state of N$_2^+$ (X$^2\Sigma_{g}^{+}$) molecule relative to X$^1\Sigma_{g}^{+}$ state of N$_2$ molecule are illustrated in Fig.~\ref{fig:IP}.

From a comparison of the PECs for the ground states of N$_2^+$ and N$_2$, see Fig.~\ref{fig:IP}, we extract the vertical ionization potential for the $\rm N_2$ molecule as equal to 15.562~eV, which corresponds to the equilibrium distance of $\rm N_2$. The discrepancy with experimental data~\cite{NIST} is 0.02 eV. Taking into account relaxation effects, the ionization potential for $\rm N_2$ molecule can be found as 15.537~eV. Within \texttt{DIRAC} program package we made both relativistic and non-relativistic calculations of PECs. As a result, we estimate the relativistic correction to the vertical ionization potential to be about 0.03 eV. This value is an order of magnitude smaller than the calculation uncertainty.

\begin{figure}
    \centering
    \includegraphics[width=0.45\textwidth]{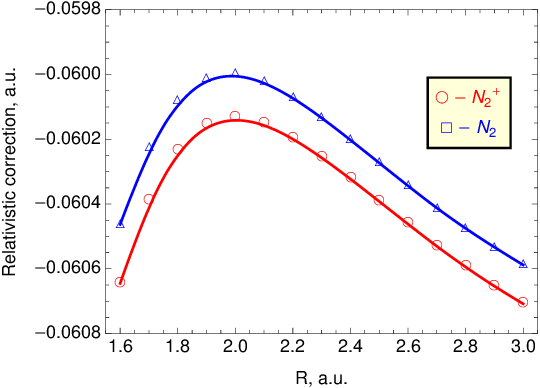}
    \caption{Relativistic corrections to the ground state total energies of N$_2^+$ (X$^2\Sigma_g^+$) and N$_2$ (X$^1\Sigma_g^+$) molecules  as a function of internuclear distance $R$. CCSD(T)/CBS values are obtained using the \texttt{DIRAC} code and given in atomic units.}
    \label{fig:REL}
\end{figure}

The comparison of relativistic and non-relativistic CCSD(T)/CBS results for both N$_2^+$ and N$_2$ systems allows us to extract relativistic corrections, which we consider as the difference between relativistic and non-relativistic results. These corrections can be described by Morse potential. The results of calculations are illustrated in Fig.~\ref{fig:REL} as a function of internuclear distance  $V^\mathrm{rel} (R)$.
The $V^\mathrm{rel} (R)$ function reaches its maximum value of $-0.0601$~a.u. at internuclear distances of $R_{\mathrm{max}} \approx 2.00$~a.u. for the N$_2^+$ molecular ion, which is slightly less than the equilibrium distance. A similar dependence of relativistic corrections on internuclear distance was found for the CO molecule \cite{konovalova18, usov24}. Note that the CCSD(T)/DPT2 approach, which accounts for the leading relativistic corrections to energy and response properties \cite{cfour}, cannot reproduce the same dependence of the relativistic correction, especially in the vicinity of small internuclear distances.

The computational results obtained using the \texttt{CFOUR} and \texttt{DIRAC} packages indicate that the inclusion of relativistic effects does not significantly influence the accuracy of theoretical predictions when compared to experimental observations of the vibrational structure. 
Specifically, from Fig.~(\ref{fig:REL}) one can see that the contribution of relativistic effects to the total energy of molecules in the considered range of inter-nuclear distances changes insignificantly, i.e., varies at the level of $10^{-7}$. Consequently, such a change should be weakly manifested in calculations of the vibrational spectra. Moreover, this analysis supports the validity of employing nonrelativistic calculations for determining the polarizabilities of the considered molecule, as discussed in the following section.

As a subsequent step, we calculate the molecular constants for the obtained potential energy curves. The numerical results for the molecules N$_2^+$ and N$_2$ are summarized in Table~\ref{tab:mc}.
\begin{table}[ht]
\caption{Molecular constants for N$^+_2$ and N$_2$ compounds found by the fully relativistic CCSD(T)/CBS method. Each pair of values is given for those calculated theoretically in the CCSD(T)/CBS limit and those obtained from experimental data~\cite{Klynning1982, NIST}. The results are labeled as CBS and Exp., respectively.}
\begin{tabular*}{\columnwidth}{@{\extracolsep{\fill}}c | l l | l l }
\hline
&\multicolumn{2}{c|}{N$^+_2$}&\multicolumn{2}{c}{N$_2$}\\
 & CBS & Exp. & CBS & Exp.\\
\hline
$R_{e}$, \AA & $1.11453$ & $1.116$  & $1.09633$ & $1.09768$ \\
$\omega_{e}$, cm$^{-1}$ & $2215.10$ & $2207.19$ & $2361.47$ & $2358.57$\\
$\omega_{e}x_{e}$, cm$^{-1}$ & $15.032$ & $16.136$ & $13.651$ & $14.324$\\
\hline
\end{tabular*}
\label{tab:mc}
\end{table}
Comparing our results with those reported in \cite{Klynning1982, NIST}, we observe good agreement. The largest discrepancy, at the level of several percent, is found only for the first anharmonicity coefficient, $\omega_{e}x_{e}$. As before, the molecular constants obtained using the CCSD(T)/CBS approach provide substantial validation of the chosen methodology.

\section{Vibrational structure calculations and thermal Stark shifts}
\label{results}

By solving equation (\ref{1}) using the previously obtained potential energy curves (PECs), we determined the vibrational energy levels for the molecular ion N$_{2}^{+}$. The results are summarized in Table~\ref{tab:vibr}. Specifically, Table~\ref{tab:vibr} includes data obtained from both theoretical calculations and experimental results previously reported \cite{Singh1966, Lofthus1977, Langhoff1987, Nagy2003, Ferchichi2022}. Our calculations pertaining to the three intermediate columns of this table demonstrate strong concordance both among themselves and with prior calculations. 

Direct comparison of the relativistic (\texttt{DIRAC}) and nonrelativistic calculations (\texttt{CFOUR}) in the CBS limit show the insignificance of the relativistic corrections, whose contribution is expected at the level of the second decimal place, see Fig.~\ref{fig:REL}. Remaining acceptable, the correspondence between experimental and theoretical values for the vibrational energies deteriorates with increasing vibrational quantum number $\nu$. 

Finally, the primary objective of this work is to accurately calculate the polarizabilities of the N$_2^+$ ion and subsequently apply these values to estimate the thermal Stark shifts for vibrational levels. Based on the above analysis, the numerical calculations of the polarizabilities were performed within \texttt{NWChem} package at the CCSD level using the same basis sets, i.e. aug-cc-PV$n$Z. The results obtained for the N$_{2}^{+}$ molecule as well as for individual atoms (see section~\ref{part1}) are collected in Table~\ref{tab:final}.
\begin{widetext}
\begin{table*}[ht]
\caption{The averaged static dipole polarizability $\alpha^{\mathrm{int}}_{\nu}(0)$ for N$^+_2$ molecule in $a_{0}^3$ units ($a_{0}$ is the Bohr radius), see Eq.~(\ref{avg}). The leading-order contribution to the BBR-induced Stark shift $\Delta f_{\nu}$ (in Hz) for different vibrational levels $\nu$ is also given for various vibrational levels $\nu \in [0,15]$. The CBS limits for polarizabilities calculated with aug-cc-PV$n$Z basis sets are established in accordance with the Eq.~(\ref{CBS_alpha}).}
\centering
\begin{tabular}{ c c c c c c c c c c c}
\hline
\textbf{$\nu$} 
& \multicolumn{1}{c}{$\alpha^{\mathrm{int}}_{\nu}$} & \multicolumn{1}{c}{$\alpha^{\mathrm{int}}_{\nu}$} & \multicolumn{1}{c}{$\alpha^{\mathrm{int}}_{\nu}$} & \multicolumn{1}{c}{$\alpha^{\mathrm{int}}_{\nu}$} & \multicolumn{1}{c}{$\alpha^{\mathrm{int}}_{\nu}$} 
& \multicolumn{1}{c}{$\Delta f_{\nu}$} & \multicolumn{1}{c}{$\Delta f_{\nu}$} & \multicolumn{1}{c}{$\Delta f_{\nu}$} & \multicolumn{1}{c}{$\Delta f_{\nu}$} & \multicolumn{1}{c}{$\Delta f_{\nu}$} \\
& \multicolumn{1}{c}{Sadlej} & \multicolumn{1}{c}{Sadlej+} & \multicolumn{1}{c}{aug-cc-PVTZ} & \multicolumn{1}{c}{aug-cc-PVQZ} & \multicolumn{1}{c}{CBS} 
& \multicolumn{1}{c}{Sadlej} & \multicolumn{1}{c}{Sadlej+} & \multicolumn{1}{c}{aug-cc-PVTZ} & \multicolumn{1}{c}{aug-cc-PVQZ} & \multicolumn{1}{c}{CBS} \\
\hline 
0  & 1.00 & 0.99 & 1.58 & 1.44 & 1.35 & -0.00137 & -0.00136 & -0.00216 & -0.00198 & -0.00185 \\
1  & 1.14 & 1.13 & 1.71 & 1.59 & 1.50 & -0.00156 & -0.00155 & -0.00234 & -0.00217 & -0.00206 \\
2  & 1.29 & 1.28 & 1.88 & 1.75 & 1.67 & -0.00176 & -0.00175 & -0.00257 & -0.00240 & -0.00228 \\
3  & 1.46 & 1.45 & 2.08 & 1.95 & 1.86 & -0.00200 & -0.00199 & -0.00285 & -0.00267 & -0.00255 \\
4  & 1.67 & 1.67 & 2.27 & 2.18 & 2.11 & -0.00229 & -0.00228 & -0.00311 & -0.00298 & -0.00290 \\
5  & 1.94 & 1.93 & 2.47 & 2.45 & 2.44 & -0.00265 & -0.00265 & -0.00339 & -0.00336 & -0.00335 \\
6  & 2.28 & 2.26 & 2.83 & 2.81 & 2.80 & -0.00311 & -0.00310 & -0.00388 & -0.00386 & -0.00384 \\
7  & 2.69 & 2.67 & 3.42 & 3.23 & 3.10 & -0.00368 & -0.00366 & -0.00469 & -0.00443 & -0.00424 \\
8  & 3.09 & 3.07 & 3.97 & 3.54 & 3.23 & -0.00423 & -0.00420 & -0.00544 & -0.00485 & -0.00442 \\
9  & 3.25 & 3.23 & 4.04 & 3.51 & 3.11 & -0.00445 & -0.00442 & -0.00551 & -0.00481 & -0.00427 \\
10 & 3.10 & 3.08 & 3.58 & 3.14 & 2.83 & -0.00425 & -0.00422 & -0.00491 & -0.00431 & -0.00388 \\
11 & 2.83 & 2.80 & 3.13 & 2.80 & 2.57 & -0.00387 & -0.00384 & -0.00429 & -0.00384 & -0.00352 \\
12 & 2.53 & 2.50 & 3.01 & 2.68 & 2.44 & -0.00346 & -0.00343 & -0.00412 & -0.00367 & -0.00334 \\
13 & 2.26 & 2.24 & 2.89 & 2.56 & 2.33 & -0.00310 & -0.00307 & -0.00396 & -0.00351 & -0.00319 \\
14 & 2.10 & 2.07 & 2.70 & 2.42 & 2.21 & -0.00287 & -0.00284 & -0.00371 & -0.00331 & -0.00303 \\
15 & 1.99 & 1.96 & 2.65 & 2.36 & 2.15 & -0.00273 & -0.00269 & -0.00363 & -0.00324 & -0.00295 \\
\hline
\end{tabular}
\label{tab:final}
\end{table*}
\end{widetext}

Extrapolation of polarizabilities towards the CBS limit is done according to an equation similar to Eq. (\ref{CBS}):
\begin{eqnarray}
\label{CBS_alpha}
    \alpha_{\mathrm{CBS}}=\frac{\alpha_{n}n^3-\alpha_{n-1}(n-1)^3}{n^3-(n-1)^3}.
\end{eqnarray}
This formula can be easily obtained from Helgaker’s two-point extrapolation formula by taking second derivatives with respect to the electric field \cite{Helgaker1997}.

As can be seen from Table~\ref{tab:final}, the total line shift $\Delta f_{\nu'\nu} = \Delta f_{\nu'}-\Delta f_{\nu}$ remains largely unaffected by the choice of basis set, with minimal variation observed.
For the lines proposed in the work by Kajita et al. \cite{Kajita2014}, the calculated blackbody radiation (BBR) Stark shifts for the transition frequencies $\nu = 0 \rightarrow \nu = 1$ and $\nu = 0 \rightarrow \nu = 2$, which are 65.2 THz and 129.4 THz respectively \cite{Gilmore1992}, are found to be $-0.00021$ Hz and $-0.00043$ Hz. These values were obtained using the CBS(3,4) limit as the final result, as presented in column 11 of Table~\ref{tab:final}. The fractional uncertainties $\Delta f_{\nu'\nu}/f_{\nu'\nu}$ associated with these shifts are $ -3.22\times 10^{-18} $ and $-3.32\times 10^{-18}$, respectively, and are consistent with the estimates provided in \cite{Kajita2014}. 

Dynamic corrections to the line shifts $\Delta f_{01}$ and $\Delta f_{02}$ are associated with terms containing an $n \ge 1$ in Eq.~(\ref{BBR}). The corresponding contribution of the next order scales with the six power of the temperature and is three orders of magnitude smaller than the shift induced by the static term. Consequently, the relative uncertainty due to the dynamic corrections is approximately of the order of $10^{-21}$. It can be noted that in \cite{Carollo2018} the comparable Stark BBR shifts were recently calculated for the ground state of the O$^{+}_2$ molecule. The static polarizabilities calculated there were obtained using an approximate technique that involves summation over transitions between vibrational states of the two nearest electronic states, and is less accurate than the approach presented here.

In addition, we carried out calculations in the specialized Sadlej-PVTZ (Sadlej) and Sadlej+  sets adapted to calculate molecular properties such as polarizability and dipole moment \cite{Sadlej_1992, Baranowska_2009}. The energy shifts found in these basis sets are summarized in Table~\ref{tab:final} for comparison with other calculations. The data found for the polarizabilities of N$^+_2$ and $N_2$ molecules, as well as for individual N and N$^+$ atoms, are collected in Table~\ref{tab:single}.
\begin{table}[ht]
\caption{The static dipole polarizabilities of individual atoms (columns 2 and 3), as well as the N$_2$ and N$_2^+$ molecules at equilibrium distance (columns 4 and 5) in units of $a_{0}^3$ (where $a_{0}$ is the Bohr radius) for various basis sets. The CBS limits for aug-cc-PV$n$Z basis sets are established in accordance with the Eq.~(\ref{CBS_alpha}).}
\centering
\begin{tabular}{ l c c c c }
\hline
 & N & N$^+$ & N$_2$ & N$_2^{+}$\\
\hline  
Sadlej & 7.98 & 3.46 & 11.87 & 12.52 \\
Sadlej+ & 8.00 & 3.49 & 11.93 & 12.56\\
aug-cc-PVTZ & 7.59 & 3.59 & 11.67 & 12.75 \\
aug-cc-PVQZ & 7.74 & 3.58 & 11.63 & 12.76\\
aug-cc-PV5Z & 7.74 & 3.56 & 11.59 & 12.77\\
aug-cc-PV6Z & 7.77 & 3.57 & -  & - \\
CBS(3,4) & 7.85 & 3.57 & 11.60 & 12.77 \\
CBS(4,5) & 7.74 & 3.54 & 11.55 & 12.78\\
CBS(5,6) & 7.81 & 3.58 & -  & -\\
NIST values & 7.80\footnote{DFT calculations with B3LYP exchange functional and Sadlej basis set \cite{NIST}} & 3.71\footnote{MP2 calculations with aug-cc-PVTZ basis set} & 11.52\footnote{CCSD calculations with aug-cc-PVQZ basis set} & 11.51\footnote{CBS(3,4) limit of DFT calculations with PBEPBE functional}\\
              & 8.28\footnote{DFT calculations with PBEPBE exchange functional and Sadlej basis set \cite{NIST}} & 3.68\footnote{CBS(4,5) limit of CCSD(T) calculations with t-aug-PVQZ and t-aug-PV5Z triply augmented basis sets \cite{_hn_2020}} & & \\
Experiment & 7.6(4) & - &  11.54(58)\footnote{See \cite{Olney1997}, where the uncertainty for presented value is estimated to be 5\%} &  - \\
\hline 
\end{tabular}
\label{tab:single}
\end{table}

\begin{figure}
    \centering
    \includegraphics[width=0.45\textwidth]{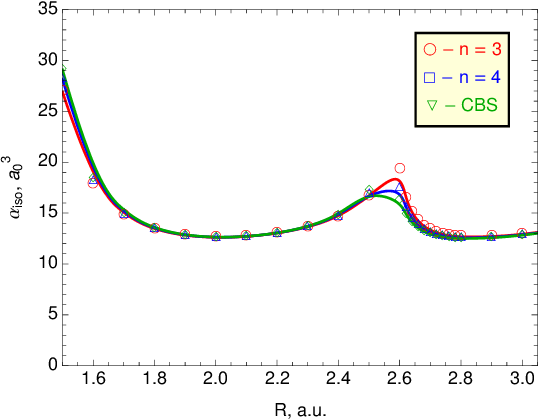}
    \caption{Isotropic part of static dipole polarizability $\alpha_{\mathrm{iso}}=[2\alpha_{xx}+\alpha_{zz}]/3$ for the ground state of  N$_2^+$ molecule calculated at CCSD level and aug-CC-PV$n$Z ($n=3,\,4$) basis sets. CBS limit is taken according to Eq.~(\ref{CBS_alpha}). All the values are in atomic units.}
    \label{fig:3}
\end{figure}

\begin{figure}
    \centering
    \includegraphics[width=0.45\textwidth]{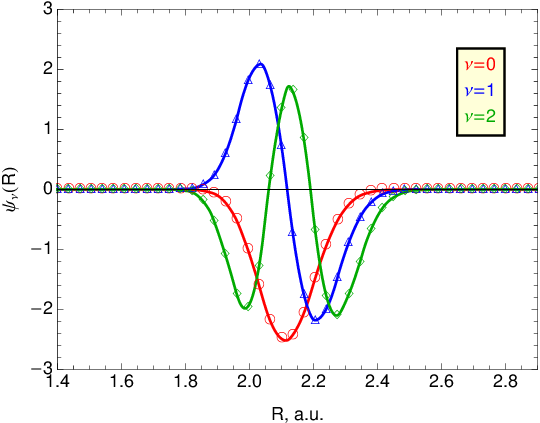}
    \caption{The plot of first three vibrational wave functions $\psi_{\nu}(R)$ ($\nu=0,\,1,\,2$) corresponding to eigenvalues of Table~\ref{tab:vibr}.}
    \label{fig:4}
\end{figure}

The behavior of the static polarizability with the variation of the internuclear distance is shown in Fig.~\ref{fig:3}.
The wave functions of the first three vibration states constructed by solving Eq. (\ref{1}) are shown in Fig.~\ref{fig:4}.

As follows from Fig.~\ref{fig:3} the polarizability peak is observed in the internuclear distance interval around $(2.5,2.7)$ a.u. The origin of this behavior is related to the close location of the ground X$^2\Sigma_g^+$ and excited A$^2\Pi_{u}$ electron terms in this region (see Fig.~1 in \cite{Nagy2003}). The latter, however, does not play a significant role in calculating the average polarizabilities according to Eq.~(\ref{avg}) and the thermal Stark shifts for low-lying states, see Eq.~(\ref{BBR}). Since the vibrational wave functions are almost zero at these distances even for excited states, see Fig.~\ref{fig:4}, the contribution of this peak is negligible.

To ensure the completeness of our calculations, we also compare different extrapolation schemes for polarizabilities, see the corresponding rows in Table~\ref{tab:single}. Additional calculations were also performed using doubly augmented correlation-consistent polarized valence $n$-tuple zeta (d-aug-cc-PV$n$Z) basis sets. No significant changes were observed in comparison with singly augmented ones. For comparison, the last row of the Table~\ref{tab:single} presents the experimentally obtained polarizability values for the N and N$_2$ systems. There is good agreement between the results.

Finally, we estimate the contributions of scalar relativistic and quantum electrodynamic (QED) effects to polarizabilities. To achieve this, we employed generalized relativistic effective core potentials (GRECPs) \cite{Titov1999, Mosyagin2013}, which incorporate QED effects non-perturbatively via a model operator now extended for molecular systems \cite{modelQED}. It is found that the relative change in dipole polarizability resulting from scalar relativistic and QED effects does not exceed 0.5\%. \\

\section{Conclusions}

In this work we have calculated thermally induced shifts to vibrational states of the molecule within the coupled cluster method. The basics of thermal shift theory are given in section~\ref{part1}, where the main steps of the necessary calculations are also given. In particular, as a first step the potential curve for the ground electronic states X$^2\Sigma_g^+$ within the CCSD(T) approach was constructed. The numerical results and the corresponding comparative analysis involving the accuracy of calculations are given in section~\ref{part0}. The obtained numerical results for PECs allowed us to accurately estimate the vibrational energy structure of the N$_2^+$ molecular ion (which we are focusing on) and the N$_2$ molecule. The details of corresponding calculations as well as the comparative analysis are given in section~\ref{results}.

As a final result, the dependencies of dipole polarizabilities on the internuclear distance, see Fig.~\ref{fig:3}, as well as the polarizabilities of individual atoms, were calculated, see Table~\ref{tab:single}. Our results significantly improves previous calculations of the vibrational structure and are in much better agreement with experimental results. This also supports the validity of using the single-reference approach within the specified range of internuclear distances.

To summarize, our numerical calculations not only confirm the previously obtained estimates but also reinforce the potential of the molecular nitrogen ion as a prime candidate for exploring variations of fundamental constants and creating a new generation of frequency standards. This positions the molecular nitrogen ion as a very promising system for developing molecular clocks capable of achieving relative errors below $10^{-18}$ \cite{Kajita2014}. Such precision holds significant implications for advancements in fundamental physics and timekeeping technology.

\section*{Acknowledgements}
Calculations of the electronic structure in section~\ref{part0} were carried out with the support of the Foundation for the Advancement of Theoretical Physics and Mathematics "BASIS" (grant No.~23-1-3-31-1). Calculations of thermal polarizabilities and thermal Stark shifts in the section~\ref{results} were carried out with the support of Russian Science Foundation under grant No.~22-12-00043.
The authors would like to express their gratitude to prof. A. V. Stolyarov (Faculty of Chemistry, Moscow State University) and prof. A. V. Titov (Petersburg Nuclear Physics Institute) for his insightful discussions and expert advices. The updated RKR potential curve employed in this study was supplied by prof. A. V. Stolyarov.

\bibliographystyle{apsrev4-1}
\bibliography{sample}

\end{document}